\documentclass[aps,prb,twocolumn,nofootinbib]{revtex4-2}

\usepackage{tikz}
\usepackage{pgfplots}
\usepackage{subcaption}
\usetikzlibrary{arrows.meta}
\usetikzlibrary{backgrounds}
\usepgfplotslibrary{patchplots}
\usepgfplotslibrary{fillbetween}
\pgfplotsset{%
layers/standard/.define layer set={%
background,axis background,axis grid,axis ticks,axis lines,axis tick labels,pre main,main,axis descriptions,axis foreground%
}{
grid style={/pgfplots/on layer=axis grid},%
tick style={/pgfplots/on layer=axis ticks},%
axis line style={/pgfplots/on layer=axis lines},%
label style={/pgfplots/on layer=axis descriptions},%
legend style={/pgfplots/on layer=axis descriptions},%
title style={/pgfplots/on layer=axis descriptions},%
colorbar style={/pgfplots/on layer=axis descriptions},%
ticklabel style={/pgfplots/on layer=axis tick labels},%
axis background@ style={/pgfplots/on layer=axis background},%
3d box foreground style={/pgfplots/on layer=axis foreground},%
},
}

\usepackage[a-1b]{pdfx}

\usepackage{dcolumn}
\usepackage{bm}
\usepackage{multirow}

\usepackage[normalem]{ulem}

\usepackage{amsmath}
\usepackage{amsthm}
\usepackage{amstext}
\usepackage{amssymb}
\usepackage{mathrsfs}
\usepackage{amsfonts}
\usepackage{amsbsy} 

\usepackage[all,matrix,cmtip]{xy}
\usepackage{color}

\usepackage{hyperref}
\hypersetup{
    colorlinks = true,
    allcolors = blue,
}

\definecolor{red}{rgb}{1,0,0}
\definecolor{blue}{rgb}{0,0,1}
\definecolor{dblue}{rgb}{0,0,0.4}
\definecolor{green}{rgb}{0,1,0}
\definecolor{black}{rgb}{0,0,0}
\definecolor{white}{rgb}{1,1,1}

\definecolor{brn}{rgb}{.8,.4,.0}
\definecolor{redo}{rgb}{1,.5,.0}
\definecolor{ddgrn}{rgb}{0,0.4,0}
\definecolor{dgrn}{rgb}{0,0.55,0}
\definecolor{dbl}{rgb}{0,0,0.5}

\usepackage[bbgreekl]{mathbbol}

\newcommand{\Z}{\mathbb{Z}}

\renewcommand{\v}[1]{\boldsymbol{#1}} 
\newcommand{\tl}[1]{\widetilde{#1}} 
\newcommand{\ii}{\hspace{1pt}\mathrm{i}\hspace{1pt}}
\newcommand{\ee}{\hspace{1pt}\mathrm{e}}
\newcommand{\dd}{\hspace{1pt}\mathrm{d}}
 
\renewcommand{\>}{\rangle} 

\newcommand{\Rf}[1]{Ref.~\onlinecite{#1}}

\newcommand{\prt}{\partial}

\newcommand{\ie}{{\it i.e.~}}

\newcommand{\bpm}{\begin{pmatrix}}
\newcommand{\epm}{\end{pmatrix}}
\newcommand{\bmm}{\begin{matrix}}
\newcommand{\emm}{\end{matrix}}

\usepackage{euscript}

\newcommand{\eps}{\epsilon}

\hypersetup{
colorlinks = true,
linkbordercolor = {white}
}
\begin{document}
\title{Variational Monte Carlo Optimization of Topological Chiral
Superconductors}

\author{Minho Luke Kim}

\author{Abigail Timmel}

\author{Xiao-Gang Wen}

\affiliation{Department of Physics, Massachusetts Institute of Technology,
Cambridge, Massachusetts 02139, USA
}

\begin{abstract}

We perform the variational Monte Carlo calculation for recently proposed chiral
superconducting states driven by strong Coulomb interactions.  We compare the
resulting energetics of these electronic phases for the electron dispersion
relation $E_k = c_2 k^2+c_4 k^4$.  Motivated by the recent discovery of chiral
superconductivity in rhombohedral graphene systems, we apply our analysis to
relevant parameter regimes. We demonstrate that topological chiral
superconducting phases (including a spin-unpolarized state) can be
energetically favored over the spin-valley polarized Fermi liquid above the
density of Wigner crystal phase.  Our results show that the preference for
chiral superconductivity is strongest when $c_2$ lies between zero and a
negative value corresponding to a Fermi sea on the verge of forming a hole pocket
around $k=0$. This finding suggests that superconductivity can arise from pure
repulsive Coulomb interactions in systems with an almost flat band bottom,
without relying on the pairing instability of a Fermi surface.  This mechanism
opens a new pathway to superconductivity beyond the conventional BCS mechanism.

\end{abstract}

\maketitle

\setcounter{tocdepth}{1}
{\small \tableofcontents }

\section{Introduction}

\begin{figure*}[t]
	\begin{subfigure}[t]{0.66\columnwidth}
		\centering
		\includegraphics[width=2.2in]{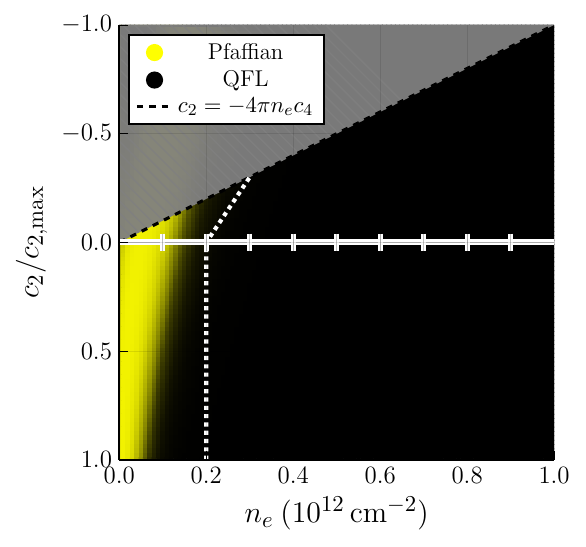}\
	\end{subfigure}
	\begin{subfigure}[t]{0.66\columnwidth}
		\centering
		\includegraphics[width=2.2in]{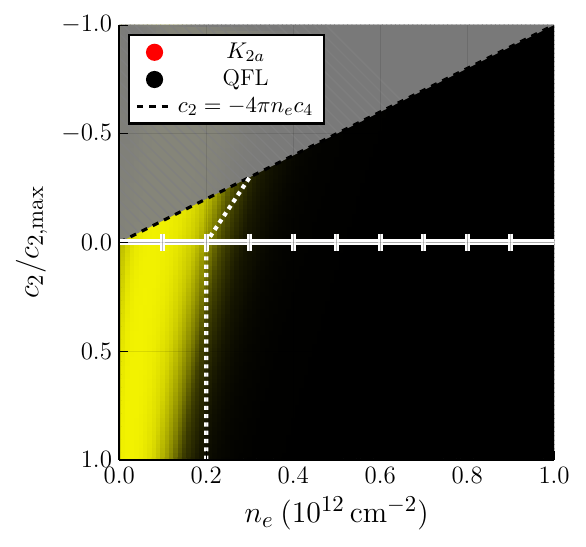}\
	\end{subfigure}
	\begin{subfigure}[t]{0.66\columnwidth}
		\centering
		\includegraphics[width=2.2in]{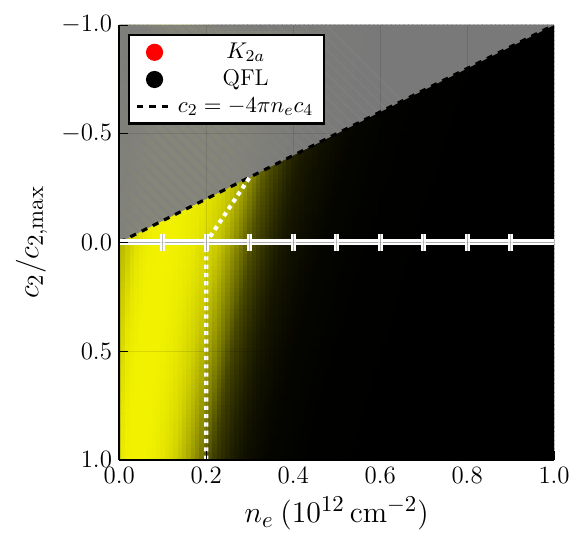}
	\end{subfigure}
	\\[-2mm]
	\begin{subfigure}[t]{0.66\columnwidth}
		\centering
		\includegraphics[width=2.2in]{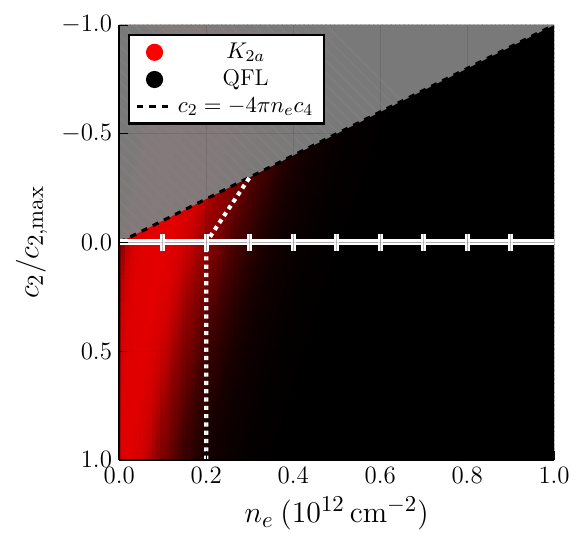}\
	\end{subfigure}
	\begin{subfigure}[t]{0.66\columnwidth}
		\centering
		\includegraphics[width=2.2in]{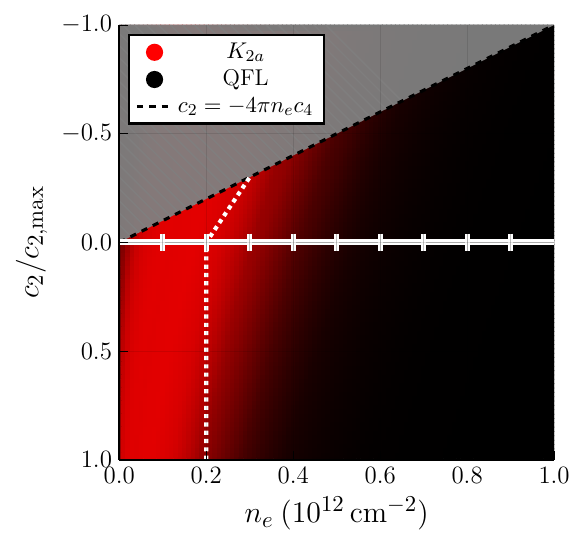}\
	\end{subfigure}
	\begin{subfigure}[t]{0.66\columnwidth}
		\centering
		\includegraphics[width=2.2in]{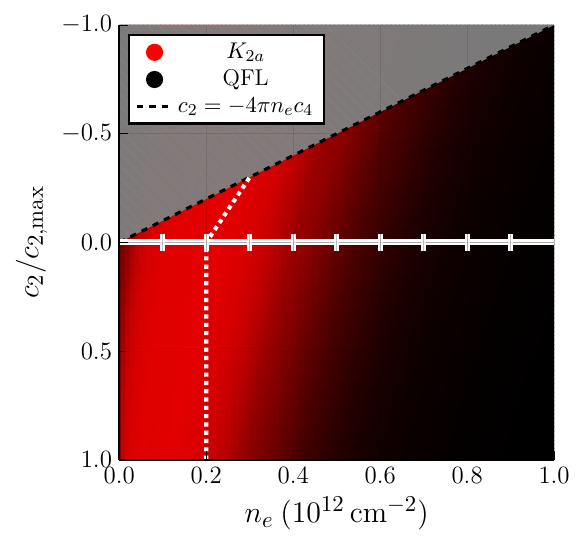}
	\end{subfigure}
	\\[-2mm]
%	\begin{subfigure}[t]{0.66\columnwidth}
%		\centering
%		\includegraphics[width=2.2in]{Figures/phaseindiv/phase_K2b_10549.pdf}\
%	\end{subfigure}
%	\begin{subfigure}[t]{0.66\columnwidth}
%		\centering
%		\includegraphics[width=2.2in]{Figures/phaseindiv/phase_K2b_5549.pdf}\
%	\end{subfigure}
%	\begin{subfigure}[t]{0.66\columnwidth}
%		\centering
%		\includegraphics[width=2.2in]{Figures/phaseindiv/phase_K2b_5366.pdf}
%	\end{subfigure}
	\begin{subfigure}[t]{0.66\columnwidth}
		\centering
		\includegraphics[width=2.2in]{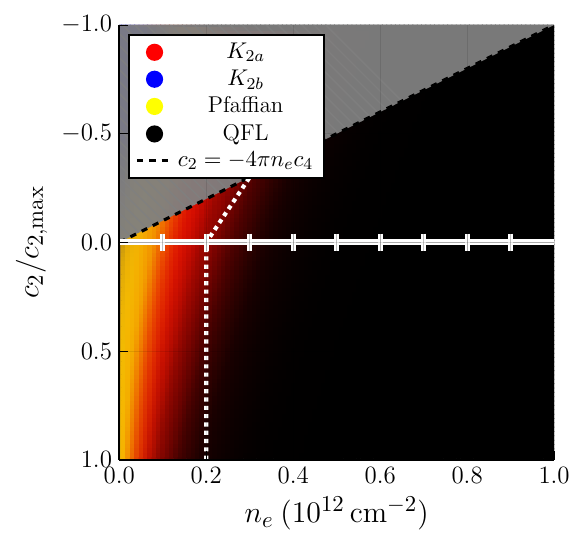}\
	\end{subfigure}
	\begin{subfigure}[t]{0.66\columnwidth}
		\centering
		\includegraphics[width=2.2in]{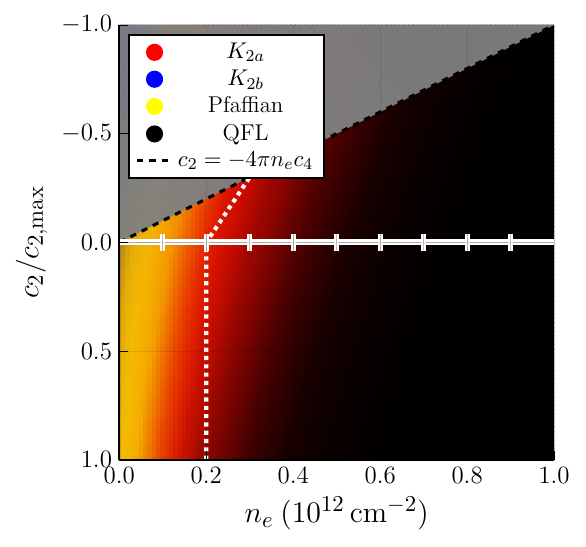}\
	\end{subfigure}
	\begin{subfigure}[t]{0.66\columnwidth}
		\centering
		\includegraphics[width=2.2in]{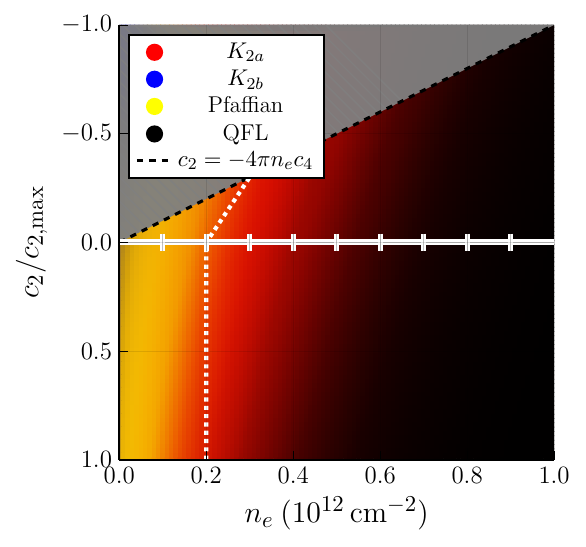}
	\end{subfigure}
	\caption{ Phase diagrams for the electron dispersion
\(E_k=c_2k^2+c_4k^4\) as functions of \(c_2\) and electron density \(n_e\). The
\textbf{first row}  shows the competition between the QFL and Pfaffian chiral
superconducting phase (blue), while the \textbf{second row} shows that between
the QFL and \(K_{2a}\) chiral superconducting phase (red). The \(K_{2b}\) phase
is always unfavorable in the relevant parameter regime. The \textbf{third row}
combines all competing phases. Experimentally, \(c_2\) can be tuned by the
displacement field. The slanted line marks the Fermi-surface transition
(Fig.~\ref{FStrans}); our calculations are valid only below this line. For the
samples of Ref.~\cite{HJ240815233}, a Wigner crystal is observed for \(n_e
\lesssim 0.2\times10^{12}\,\mathrm{cm}^{-2}\) at \(c_2=0\), and for \(n_e
\lesssim 0.3\times10^{12}\,\mathrm{cm}^{-2}\) at the Fermi-surface transition,
as indicated by the dotted white line. The color intensity is determined from
the Boltzmann weight $\frac{e^{-E_p/\Delta E}}{\sum_p e^{-E_p/\Delta E}}$,
where \(E_p\) is the energy of phase \(p\) and \(\Delta E\) is the numerical
uncertainty. The three columns correspond to: \textbf{left}, \(\epsilon=10\),
\(c_4=549\,\mathrm{meV\,nm}^4\), \(c_2\in[69,-69]\,\mathrm{meV\,nm}^2\);
\textbf{middle}, \(\epsilon=5\), \(c_4=549\,\mathrm{meV\,nm}^4\),
\(c_2\in[69,-69]\,\mathrm{meV\,nm}^2\); \textbf{right}, \(\epsilon=5\),
\(c_4=366\,\mathrm{meV\,nm}^4\), \(c_2\in[46,-46]\,\mathrm{meV\,nm}^2\). The
parameters are chosen to be representative of four-layer
graphene~\cite{HJ240815233}.  } \label{fig:phase}
\end{figure*}
\begin{figure}[t]
	\centering
	\includegraphics[width=1.5in]{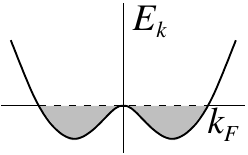}
	\caption{Electron dispersion $E_k = c_2 k^2 + c_4 k^4$ has a Fermi surface
		transition when $c_2$ satisfies $c_2 k_F^2 + c_4 k_F^4 = 0$ and the Fermi
		momentum $k_F$ is determined by the electron density $n_e$.  At the Fermi
		surface transition, a hole pocket is generated at $k=0$ as we make $c_2$ more
		negative.  } \label{FStrans}
\end{figure}

Tunable 2D electron systems have been demonstrated to be rich systems harboring
various phases of matter derived from strong correlations and topology. One of
the most striking features of physics discovered in these systems is the
emergence of superconductivity in flat band materials, pioneered by discoveries
in Moir\'e systems~\cite{BDEY2016, NY24460, AM2019} and
subsequently in rhombohedral-stacked graphene systems. For the latter,
superconductivity has now been observed in bilayer~\cite{ZY22375, ZN23613,
HY2521, LL24631, ZN25641}, trilayer~\cite{ZY21598, YJ240809906, PY240810190},
tetra and pentalayer~\cite{HJ240815233, PY25641}, and recently in hexalayer
systems~\cite{ML250405129}, pointing towards the existence of
chiral (\ie time-reversal and reflection symmetry breaking) superconducting
states between the spin-valley-polarized Fermi liquid state, also known as
quarter Fermi liquid (QFL), and Wigner crystal state. 

This unusual observation of chiral superconductivity has motivated a flurry of
theoretical work which seeks to explain the superconductivity by various
means~\cite{CS240906701, GF240913829, KW240918067, JL241109664, SW241108969,
YZ241102503, QW241207145, YXBZ250217555, PG250219474, SS250312650,
CSS250316391, CBS250315471}.  Many of these theories suggest topological chiral
superconductivities with chiral edge states. Experimentally, the topological
character and chiral edge states of those superconducting regions remain open
questions, along with the relationship of the superconductivity with its
neighboring phases. Since both the Wigner crystal phase and quarter Fermi
liquid are induced by a strong repulsive Coulomb interaction, we postulate
that the observed chiral superconductivity may also
be induced by a repulsive Coulomb interaction. The superconductor also has a short coherence length similar to electron
separation, which implies that it may not be a weak
BCS superconductor.

Recent experiment also showed the chiral superconductivity to be robust against
5T magnetic fields parallel to the plane \cite{HJ240815233}. This raises a
possibility that the observed superconductivity may be fully
spin-valley polarized. Many theoretical works, such as \Rf{CS240906701,
GF240913829}, approach the chiral superconductivity via the instability of
fully spin-valley polarized Fermi surface, \ie the BCS
mechanism. 
For example, at electron densities \(n_e>0.3\times10^{12}\,\mathrm{cm}^{-2}\), Ref.~\cite{GF240913829} proposed a phase diagram in which a non-topological spinless \(p+\ii p\) BCS superconductor appears for a ring-like Fermi surface. 
This regime corresponds to the high-displacement-field region above the dashed Fermi-surface transition line in Fig.~\ref{fig:phase}. A much weaker topological spinless \(p+\ii p\) BCS superconductor may also occur for disk-like Fermi surfaces in the same density range, corresponding to the low-displacement-field region below the dashed line in Fig.~\ref{fig:phase}; see Fig.~\ref{SCphaseFu}.

These results suggest that the BCS-type Fermi-surface instability is much stronger for a ring-like Fermi surface than for a disk-like one, due to enhanced small-wave-vector screening that helps generate an attractive \(p\)-wave channel~\cite{GB21127,GF240913829}. Experimentally, however, superconductivity is stronger at low displacement fields, where the Fermi surface is disk-like, than at high displacement fields, where it is ring-like. This discrepancy suggests that the stronger superconductivity observed at low displacement fields may arise from a mechanism distinct from the conventional BCS pairing instability. Here we explore such non-BCS mechanisms for superconductivity.

In this work, we perform an energetics analysis using the variational Monte
Carlo (VMC) method on the states proposed in \Rf{KW240918067}. These states are
driven by strong Coulomb interactions, similar to those in the neighboring
Wigner crystal and QFL phases. One of the proposed states is a strongly
correlated spinless $p+\ii p$ topological superconductor. If such a
superconductor is observed experimentally, we propose that it should be
understood as a strongly correlated superconductor, which we refer to as a
Pfaffian superconductor.

The idea of superconductivity emerging from purely
repulsive interactions has been explored in other works, for example in the context of the Kohn-Luttinger mechanism for BCS pairing instability~\cite{KL6515} in graphene systems~\cite{JL241109664, CS240906701, PG2305} and more generally~\cite{RK1183}, or via a beyond BCS
mechanism of anyon
condensation~\cite{L8825,CWW8901,F9070,L9137,WZ9174,W9243,Lc9902287,KLW0902,TW1328} where the Fermi surface is not so consequential.
%~\cite{CWW8901,L9137,WZ9174,W9243,Lc9902287,KLW0902,TW1328}. 
In our model, however, we posit that pure repulsive Coulomb interactions between
electrons may induce strongly correlated states without the initial presence of
anyons.  The mechanism is flux attachment, which leads to a class of trial wave
functions similar to those for fractional quantum Hall states.  In certain
attachment configurations, flux and particle density complement each other
within a single channel to support gapless density fluctuations.  This mode has
the property of being superfluid, and if carrying nonzero charge,
superconducting.  These states also have the features of short coherence length
and time-reversal symmetry (TRS) breaking, as well as non-trivial topological
order~\cite{W8987, W9039}.

Whether these states can be energetically favorable against the spin-valley
polarized Fermi liquid at densities relevant to experiments remains to be seen.
%The flux attachment ansatz naturally favors Coulomb repulsion, and the
%presence of a superfluid mode is equivalent to a condition that total angular
%momentum does not grow faster than linear in particle number. This second
%property brings the kinetic energy to the same order as that of a Fermi liquid,
%in contrast to arbitrary flux attachment where angular momentum is generally
%quadratic in particle number.  However, it is nontrivial to further
%characterize the spread of such states in momentum space, making comparison to
%a Fermi liquid difficult.  
While Ref.~\cite{KW240918067} finds that a Laughlin
wavefunction-based ansatz for the chiral superconductor can win over the
Hartree-Fock energy of the quarter Fermi liquid, it does not compare the
energetics of a fully optimized wavefunction for both the chiral
superconducting states and the Fermi liquid.

In this work, we pinpoint our attention to systems with
one and two species of electrons, guided by experimental evidence that the
superconductors observed are very likely to be valley-polarized. We will mainly
consider three types of topological chiral superconductors,  which are
the Pfaffian state for one species (\ie spin polarized), as well as $K_{2a}$
and $K_{2b}$ states for the two species (\ie spin unpolarized).  The Pfaffian
state  is in the same phase as a spinless $p+\ii p$ BCS
superconductor~\cite{RG0067}, hosting a single Majorana-fermion edge mode for
central charge $c=1/2$.  The $K_{2a}$-superconductor is in the same phase as a
spin-triplet $p+\ii p$ BCS superconductor for the spin unpolarized Fermi liquid,
hosting a single complex-fermion edge mode for central charge $c=1$.  The
$K_{2b}$-superconductor carries excitations with fractional statistics, and is
in a different phase than any BCS superconductor.

To make a more precise comparison of ground state energies of those states, we
concentrate on estimating the ground state energy of spin-valley polarized
Fermi liquid by constructing a Slater-Jastrow wavefunction for the correlated
Fermi liquid subject to a general $E_k = c_2 k^2 +c_4 k^4$ dispersion, where the result beyond the purely quadratic dispersion~\cite{C7818, TC8905} was not yet discussed.  

We also optimize the ground state energies of Pfaffian, $K_{2a}$, and $K_{2b}$
states by continuously tuning underdetermined parameters of the trial
wavefunctions.  However, such optimizations are quite crude. Although this can
optimize the interaction energy for short-range interactions, at long distances
the Pfaffian, $K_{2a}$, and $K_{2b}$ trial wavefunctions do not incorporate the
density fluctuations of the compressible superfluid mode.  Also, the effect of
Berry curvature of the band is not included, which can further lower the energy
of time-reversal symmetry breaking states \cite{JL241109664,MHD250305697},
which are the Pfaffian, $K_{2a}$, and $K_{2b}$ states.  We expect the Berry
curvature will lower the energy of $K_{2b}$ state most, since each electron
carries an average orbital angular momentum $3/2$.  The Berry curvature will
lower the energy of Pfaffian and $K_{2a}$ states less, since each electron
carries an average orbital angular momentum $1/2$.  It will lower the energy of
the Fermi liquid the least, since beyond the sublattice scale this state
features only plane waves with no macroscopic orbital angular momentum.
We also ignore trigonal warping~\cite{HJ240815233,GF240913829}, since
our mechanism is not BCS pairing indtability of Fermi surface and does not
depend on Fermi surface topology.

Despite these shortcomings, we find that the single-species Pfaffian and
two-species $K_{2a}$ chiral superconducting trial wavefunctions can still be
energetically favorable at densities as large as $0.5 \times 10^{12} cm^{-2}$
(see Figs. \ref{fig:phase}).  Experimentally,
superconductivity was observed for densities $0.2 \sim 0.7 \times 10^{12}
cm^{-2}$, near the Fermi-surface transition line (see Fig. \ref{FStrans}).
This finding shows that superconductivity can arise from pure repulsive Coulomb
interactions in systems with a flat band bottom.  This opens a new route to
superconductivity beyond the BCS mechanism (\ie without relying on the pairing
instability of a Fermi surface).

Our calculation shows that the spin-valley polarized Pfaffian state can give way to a spin-valley polarized Fermi liquid at relevant densities, and at lower density regions could become the most energetically favorable among all states considered. For this, we have designed a new ansatz which describes the Pfaffian state which is more energetically favorable than its Laughlin counterpart.
%Since our Pfaffian wave-function is least optimized, we should not rule out the Pfaffian state at this time.

\section{Topological Chiral Superconductivity}

The wavefunction for the repulsive interaction driven superconducting state can be inspired from the Laughlin wavefunction, where the wavefunction contains high
order of zero as two electrons approach each other. In quantum Hall states,
the kinetic term is ignored due to the fact that the band is totally degenerate
within a Landau level; hence, the minimization of the energy is achieved by high
order of zeros, driven by strong repulsive interactions.

Imposing a Laughlin wavefunction in a dispersive band rather than a Landau
level presents a problem since each electron represented by the Laughlin
wavefunction carries a large momentum of order $\sqrt{N_e n_e}$, where $N_e$
the total electron number and $n_e$ is the electron density.  We can avoid this
large-momentum problem by generalizing the Laughlin wavefunction to include
antiholomorphic factors, introducing opposite fluxes to cancel some of the
phase oscillation.  These factors are $(z_i^*-z_j^*)$, and the trial
wavefunction becomes in full generality,
\begin{align}
\label{PsiKpm}
\Psi(z_i^I)
&=
\ee^{-\frac{ \sum_{i,I} |z_i^I|^2}{4l_I^2}}
\prod_{i<j,I} (z_i^I - z_j^I)^{K^+_{II}}  
\prod_{i,j,I<J} (z_i^I - z_j^J)^{K^+_{IJ}}  
\nonumber\\
&\ \ \ \ 
\prod_{i<j,I} (z_i^{I *} - z_j^{I *})^{K^-_{II}}  
\prod_{i,j,I<J} (z_i^{I *} - z_j^{J *})^{K^-_{IJ}} , 
\end{align}
where $K^\pm_{IJ}$ and $l_I$ are variational parameters.  Here $I,J$ label
different species of electrons (which are spin-valley quantum numbers in our
case), and $ z_i^I = x_i^I +\ii y_i^I $ is the coordinate of the $i^\text{th}$
electron of the $I^\text{th}$ species.

The variational parameters should satisfy some constraints. First, all elements of $K^+$ and $K^-$ should be positive to enforce zeros when electrons coincide. Second, $ K = K^+ -K^-$ is required to be a symmetric integer matrix with odd diagonal elements, to ensure that the wavefunction is both anti-symmetric and single-valued. This can be represented as:
\begin{align}
\label{Constraint}
&K^+_{IJ} = K^+_{JI} \geq 0,\ \ \ K^-_{IJ} =  K^-_{JI} \geq 0, 
\nonumber\\
&K \equiv K^+ - K^- \in \Z, \ \ \ K_{II} \text{ mod } 2 = 1
.
\end{align}
For convenience we will define $\bar{K} = K^+ + K^-$; $\bar{K}$
corresponds to the power of $\lvert z_i^I - z_j^J \rvert$ excluding the phase
term, and it is not topologically protected.

In order for an electron to have a finite momentum in $N_e \to \infty$ limit,
$K_{IJ}$ must have a zero eigenvalue with a positive eigenvector:
\begin{align}
\label{Kf}
\sum_J K_{IJ} f_J = 0 \ \ \text{ for } f_I > 0, \ \ \sum f_I = 1.
\end{align}
This is because the total angular momentum of the trial wavefunction
\eqref{PsiKpm} can be estimated by
\begin{align}
& \ \ \ \
\sum_{I} \frac{N_I(N_I-1)}{2}   K_{II}   + \sum_{I<J} N_I N_J K_{IJ}  
\nonumber \\
& = \frac12 \sum_{I,J} N_I  K_{IJ} N_J - \frac12 \sum_{I} N_I  K_{II} 
\end{align}
where $N_I$ is the number of species-$I$ electrons.  This is a measure of the
net phase winding seen by each particle from each other particle.  The
quadratic term on $N_I$, if present, causes the momentum and the kinetic energy
per electron to diverge in the thermodynamic limit $N_e = \sum_I N_I \to
\infty$.  Therefore, for this state to be viable, the quadratic energy term
must be zero, meaning $K$ must satisfy \eqref{Kf}, where $f_I = N_I/N_e$ is the
proportion of species $I$.  A more mathematically rigorous form of the kinetic
energy is presented Appendix A of \Rf{KW240918067}, from which the same constraints
are retrieved.  

%As a final comment, after cancelling out the quadratic term, the average angular momentum per electron is
%\begin{align}
%\<L\> = - \frac12 \sum_{I} f_I  K_{II} 
%\end{align}
%which is a topological invariant of the chiral superconductor.

If the wavefunction \eqref{PsiKpm} does describe a ground state, what are its
properties?  The density ratios were fixed by the kinetic energy
condition~\eqref{Kf}, but the cofluctuation in these fixed ratios is
unconstrained, leading to a gapless mode~\cite{WZ9040,KW240918067}. Such a
gapless electron density mode implies superfluidity \cite{WZ9174}, provided
that there are no other gapless excitations. To answer this question more
formally, we can consider its effective Chern-Simons field theory developed for
multi-layer FQH states; see \Rf{W9505} and \Rf{KW240918067} for a detailed
discussion.

%\begin{equation}
%\mathcal{L} = \frac{K_{IJ}}{4\pi} a_{I\mu} \partial_\nu a_{J\lambda} \epsilon^{\mu \nu \lambda} - \frac{e_I}{2\pi} A_\mu \partial_\nu a_{I\lambda} - a_{I0} l_I(x)
%\end{equation}
%where $e_I = (1, 1, ...)^T$ are the electric charges of the excitations and
%%l_I(x)$ is the $U(1)$ charge of the compact Chern-Simons theory. Combined with
%the Maxwell term, the Chern-Simons term leads to a gap in $a_{I\mu}$
%excitations invertible $K$-matrix.  When $K$ has a single zero eigenvalue, the
%effective Chern-Simons field theory has a 

Our goal in this paper is to study the viability of the chiral superconducting
states in a realistic Hamiltonian. For this, we have set the dispersion
relation to be
\begin{equation}
E_k = c_2 k^2 + c_4 k^4.
\label{eq:disp}
\end{equation}
Applying this to our wavefunction ansatz, we find the total energy to be
\begin{align}
\label{Etot}
E_\text{tot} &= \frac{e^2 \sqrt{n_e}}{\epsilon} V + c_2 \<k^2\> + c_4 \<k^4\> ,
\nonumber\\
V&\equiv 
\sum_{IJ} f_I f_J V_{IJ} 
\nonumber \\
V_{IJ} &\equiv \int
\dd^2 z \ \frac{ \sqrt{n_e}}{ 2|z|} (g_{IJ}(z)-1) .
\end{align}
are the parameters associated with the potential energy, where $g_{IJ}(z)$ is
the electron pair distribution function, computed numerically.
The dimensionless averages $\<k^2\>/n_e$, $\<k^4\>/n_e^2$ are also computed
numerically via Monte Carlo method.

\subsection{Single-species Pfaffian Wavefunction}

The restrictions given in equations~\eqref{Constraint} and \eqref{Kf} cannot be applied to single-species electron system. However, in this case we have another class of chiral superconductor obtained by utilizing the Pfaffian. 

%\begin{figure}[t]
%	\centering
%	\includegraphics[width=3.0in]{Figures/Jr}
%	\caption{A plot of $J(r)$ (blue) and $J_1(r)$ (red), 
%with $m=3$, $p=0.7$, $\xi_1=1$, $\xi_2=2$, $C=0.25$.} \label{Jr}
%\end{figure}

An ansatz we use is given as
\begin{align}
\label{PsiPf1}
\psi_{Pf} &= \textrm{Pf} \left(\frac{1}{z_i-z_j} \right)  \prod_i \left[\frac{1}{1+ \exp (\sigma (\lvert z_i \rvert - R))} \right]^s 
\nonumber\\
&\ \ \ \
\times \prod_{i<j} J(|z_i - z_j|)
\nonumber\\
J(r) &=
r^m \sqrt{ \frac{\xi_1^{2p} + r^{2p}}{\xi_2^{2(m+p)} + r^{2(m+p)}}} 
%J_1(r) &= \frac{r^m}{\xi_1^m+r^m}\Big(1+ \frac{Cr}{\xi_2 (1+\frac{r^2}{\xi_2^2})^{p} }\Big)
\end{align}
where the wavefunction is written down at $n_e=1$ units, and $R = \sqrt{N/\pi}$ is the radius of the droplet. The form of $J(r)$ is chosen for the following reasons. To ensure that the wavefunction maintains constant density in the bulk, the large-distance limit should be a constant (in this case 1). We set the short distance behavior as $r^m$ for the parameter $m$ present in a traditional Pfaffian wavefunction. This leaves the intermediate distance behavior to be determined. To lower the interaction energy, one can engineer larger interparticle spacing by concentrating weight in a small hump at the edge of a corrrespondingly broadened Paul exclusion well.  This Jastrow factor provides several parameters for optimizing such behavior.

For the energetics, there was no statistically significant deviation in total energy for $\sigma$ between 2 and 5 and $s$ between 1 and 3. This is expected as these only serve as the decay function at the edge of the droplet and hence do not influence the bulk physics. For our results, we used $\sigma=3$ and $s=2.5$. We sampled for $\xi_1$ from $0.4$ to $1.8$, $\xi_2$ from $\xi_1$ to 1.8, both in $0.2$ intervals. We sampled $m$ from 2 to 5\ in 0.2 intervals.

One can compare this with the more traditional Laughlin-related ansatz
\begin{equation}
	\psi_{Pf, alt} =  \textrm{Pf}\left(\frac{1}{z_i - z_j}\right)
	\prod_{i<j,I} \lvert z_i - z_j \rvert^m \ee^{-\frac{ \sum_{i,I} |z_i^I|^2}{4l_I^2}}.
	\label{PsiPf}
\end{equation}
The factors multiplying $\textrm{Pf}\left(\frac{1}{z_i - z_j}\right)$
essentially describe a correlated superfluid of bosons under a repulsive
interaction.  \eqref{PsiPf1} is a better trial wavefunction, since it allows greater flexibility from
algebraic long-range density correlations. From a variational calculation standpoint as well, the Laughlin-related ansatz only offers one optimization parameter, which is not ideal for capturing all the correlation effects. We retain the same order of zero $m$ as particles approach each other, and the wavefunction remains holomorphic within the Pfaffian part, so our new ansatz and the Laughlin-related ansatz will be described by the same low-energy effective field theory and therefore be related by smooth deformations. While the new ansatz can be favorable over the quarter Fermi liquid at low densities, the simpler ansatz~\eqref{PsiPf} was unfavorable to the Fermi liquid at all densities, albeit at a small margin of around 2-3\% of Coulomb energy at the best performing density

We comment that although recent proposals of intravalley pairing~\cite{CBS250315471} predict the
possibility of pairing at finite momentum $\boldsymbol{q}$ with respect to the valley center, 
leading to symmetry-breaking and pair density waves on the lattice scale, we operate within a simple 
rotationally-symmetric valley-projected theory.  All pairing occurs with center-of-mass momentum $2K$, 
which maps to the lattice with a global commensurate phase oscillation which does not break symmetry, 
induce pair density waves, or contribute to the energetics.

\section{Ground State Energy of Chiral Superconducting States}

\subsection{$K_{2a}$ and $K_{2b}$ states with two species}

From the discussion above, we know that the density ratios of different species
of particles are fixed, while the total density obtained by scaling this ratio
may fluctuate as a gapless mode.  In the trial wavefunction~\eqref{PsiKpm}, the
density is determined by both the $l_I$ and the matrix elements of $\bar K$,
which scale the exponential decay length and orders of zero respectively.  This
provides many more degrees of freedom than necessary to fix the density ratios,
so we treat them as variational parameters which it is our goal to optimize.
The parameters $l_I$, being in 1-to-1 correspondence with particle species, are
well-suited to account for the density ratios and overall scaling.  This leaves
just the elements of $\bar K$ to be optimized.  A general 2-by-2 $\bar K$ can
be written
\begin{equation}
\bar{K} = K^+ + K^- =  \begin{pmatrix}
a & c \\ c & b
\end{pmatrix}.
\label{eq:barK}
\end{equation}

The topological character of the state is encoded in the $K$-matrix, which we constrain to have a single zero eigenvector with all-positive entries as argued above.  For two species, $K$ can only take the form

\begin{equation}
K = \begin{pmatrix}
m & -m \\ -m & m.
\end{pmatrix}
\end{equation}
where $m$ must be an odd integer.  These states all have $f_I = 1/2$.  For our purposes, we will focus on $m=1$, which we will label as $K_{2a}$ and $m=3$, $K_{2b}$. Other higher valued $m$s are possible, but their more nontrivial topological order will make them less likely to appear in real-experiment settings.We note that $K_{2a}$ and $K_{2b}$ wavefunctions do not incorporate long-distance density fluctuation, which the single-species Pfaffian wavefunction does incorporate, so our multi-species energy estimates will more conservative.

%It follows that the species density ratio is given as $f_I = 1/2$ for all $I$ in the two-species case. The topological order of the superconductor is encoded in $K$, and by extension $m$. Given a chiral superconductor with a particular $K$, all variational degrees of freedom are encoded in $\bar{K}$, and by extension, its matrix elements $a$, $b$, and $c$. For our purposes, we will focus on $m=1$, which we will label as $K_{2a}$ and $m=3$, $K_{2b}$. Other higher valued $m$s are possible, but their more nontrivial topological order will make them more unlikely to appear in real-experiment settings.

For kinetic energy, we numerically calculate the dimensionless
values of $\langle k^2 \rangle/n_e $ and $\langle k^4 \rangle/n_e^2$, for $a,
b, c$ satisfying $5 \geq a = b \geq c \geq m$ in $0.2$ intervals.  
Given an ansatz wavefunction, $\langle k^2 \rangle/n_e $ and $\langle k^4 \rangle/n_e^4$ are computed using numerical derivatives of the wavefunction for electrons sufficiently inside the bulk.  Each run consists of $5 \times 10^5$ samples
with 200 electrons. To ensure independent sampling, we collect data every 200
steps, so that on average all the electrons would have had a chance to be
updated. 

The potential energy $V$ does not depend on $K$, so the same result
applies for both $K_{2a}$ and $K_{2b}$. Here, without loss of generality, we
set $5 \geq a = b \geq c \geq m$ in $0.2$ intervals for 200 electrons,
collecting $3 \times 10^6$ samples per MC run with 20 independent
configurations. We proceed by sampling the interparticle distance to numerically extract the pair distribution function, then integrate the pair distribution function with the potential energy (subjected to uniform positive charge background to maintain charge neutrality) to extract the potential energy per particle. 

For each choice of $c_2$, $c_4$ and $n_e$, we compute the total energy
$E_\text{tot}$ in \eqref{Etot} for the optimized values of $a,b,c$'s. Refer to Appendix~\ref{sec:appB} for details. %We choose $a,b,c$ that minimize the total energy to obtain the approximate ground
%state energy for $K_{2a}$ and $K_{2b}$ states.

\subsection{Pfaffian type superconductors with one species} 

Using the new Pfaffian ansatz, we have four variational parameters $m$, $p$, $\xi_1$, and $\xi_2$. We choose $2 \leq m \leq 5$, $0 \leq p \leq 5-m$ in $0.2$ intervals (so that the maximum order of zeros is $5$. ), and $0 \leq \xi_1 \leq \xi_2 \leq 1.6$ in $0.2$ intervals. For each wavefunction ansatz, we perform a numerical
derivative of the wavefunction and sample $\frac{\nabla^2
\psi_{pf}}{\psi_{pf}}$ and $\frac{\nabla^4 \psi_{pf}}{\psi_{pf}}$ for particles
sufficiently inside the droplet. While these are more costly than the
two-species case due to the computation of the Pfaffian, it is still numerically achievable to
perform VMC sampling of 70 electrons for the kinetic energy and collect $5
\times 10^5$ samples in reasonable computational time. For the potential
energy, the procedure is exactly the same as the two-species case above.

%We also attempted the Pfaffian trial wavefunction~\eqref{PsiPf}, whose only free parameter is the exponent $m$. We chose $1 \leq m \leq 5$ in $0.2$ intervals. This ansatz was energetically close to the QFL to around 1-2\% of Coulomb energy, but was not enough to overcome the QFL.

\section{Ground State Energy of the quarter Fermi liquid for General Dispersion}

For the Fermi liquid with strong Coulomb interaction, we can now go beyond the
Hartree-Fock level energy. For the quadratic dispersion, we can directly quote
the results from \Rf{TC8905}.  %In our case, the QFL is spin and valley polarized, so we use the fully polarized case.
%The Hartree-Fock energy (per electron) for a general quadratic and quartic dispersion is given in terms of density as
%\begin{equation}
%E_{HF} = c_2(2\pi n_e) + \frac{4}{3} c_4  (2\pi n_e)^2 - \frac{8}{3\sqrt{\pi}} \frac{e^2 \sqrt{n_e}}{\epsilon}
%\end{equation}
%The optimized QFL energy for the fully polarized case with $c_4=0$ was computed in \Rf{TC8905},
%\begin{align}
%\label{EQFL0}
%E_\text{QFL0} &= E_{HF} + \left[ \frac{a_0 (1+a_1 x)}{1 + a_1 x + a_2 x^2 + a_3 x^3 } \right] \frac{e^2}{2a_{B}} 
%\\
%&= \left[\frac{2}{r_s^2} -\frac{16}{3\pi r_s} +  a_0 \frac{1+a_1 x}{1 + a_1 x + a_2 x^2 + a_3 x^3 }\right] \frac{e^2}{2a_{B}}.
%\nonumber 
%\end{align}
%where $a_{B} = \frac{\epsilon}{e^2} 2c_2$, $r_s = \sqrt{\frac{1}{\pi a_B^2 n_e}}$, and $x = \sqrt{r_s}$. For the QFL (\ie spin-valley polarized Fermi liquid), the fitting parameters $a_n$ are
%\begin{equation}
%a_0 = -0.0515, \: a_1 = 340.5813, \: a_2 = 75.2293, \: a_3 = 37.0170
%\end{equation}

There is no calculation available in the literature for more general dispersion
$E_k=c_2 k^2+c_4 k^4$, so we proceed to find this numerically.  We put
the electron system on a torus where there is no boundary. To go beyond
Hartree-Fock, we employ the Slater-Jastrow wavefunction of form
\begin{equation}
\psi(R) = D(R) e^{J(R)}
\end{equation}
where $R$ is the collective coordinates $(r_1, r_2, \ldots, r_N)$, $D(R)$ is
the Slater determinant (\textit{i.e.} the Hartree-Fock wavefunction), and $J(R)
= -\sum_{i < j} f(\lvert r_i - r_j \rvert) $ is the Jastrow factor. We are only
interested in the energetics, where most of the correlation energy comes from
reducing the Coulomb repulsion when two particles approach each other. We take the
Jastrow factor to be
\begin{equation}
\label{fAB}
f(r) = A e^{-\frac{r}{B}}\left(1 + \frac{r}{B} + \frac{r^2}{2B^2} \right)
\end{equation}
where $A$ and $B$ are free variational parameters. The kinetic energy then can be extracted using numerical derivatives\footnote{We note an interesting observation that when the density profile is uniform, sampling $\frac{\nabla^4 \psi}{\psi}$ and $\left \lvert\frac{\nabla^2 \psi}{\psi} \right \rvert^2$ retrieves the same results. }. The full procedure is in Appendix~\ref{sec:appA}.

%The quadratic part of the kinetic energy can be calculated by sampling the local value $\frac{\nabla^2 \psi}{\psi}$, where an explicit formula is analytically known. The quartic part of the energy however is difficult to obtain analytically if we calculate $\frac{\nabla^4 \psi}{\psi}$, but utilizing that the torus does not have a boundary, we can sample  $\left \lvert\frac{\nabla^2 \psi}{\psi} \right \rvert^2$ instead. 

\begin{figure}[t]
\centering
\includegraphics[width=3.2in]{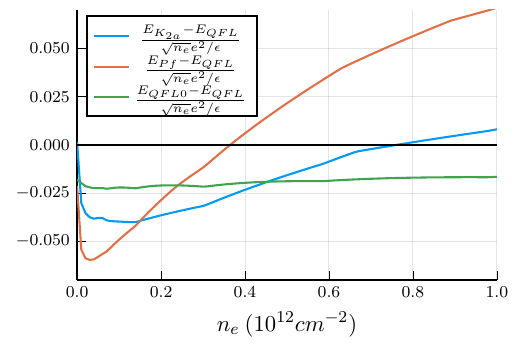}
\caption{Comparison of two ground state energies (per electron), $E_\text{QFL}$
and $E_\text{QFL0}$, for QFL.  $E_\text{QFL}$ is from our numerical calculation
and $E_\text{QFL0}$ is from \Rf{TC8905}. 
%We also plotted $\frac{E_{K_{2a}}-E_\text{QFL}}{\sqrt{n_e}e^2/\eps}$ and $\frac{E_{Pf}-E_\text{QFL}}{\sqrt{n_e}e^2/\eps}$, where $E_{K_{2a}}$ is the ground state energy (per electron) of $K_{2a}$ state and $E_{Pf}$ is the same for the Pfaffian state. 
%$\frac{E_\text{QFL0}-E_\text{QFL}}{\sqrt{n_e}e^2/\eps}$ is plotted.  
$E_{K_{2a}}$ is the ground state energy (per electron) of $K_{2a}$ state and $E_{Pf}$ is the same for the Pfaffian state.  We used the four-layer graphene parameters $c_2 = 70$ m$eV$ n$m^2$ and $\eps=5$.   }
\label{energyPlotc2}
\end{figure}

After computing $\<k^2\>$, $\<k^4\>$, and average Coulomb energy per electron
$V\sqrt{n_e}e^2/\eps$, we obtain the ground state energy $E_\text{QFL}$ (per
electron) from \eqref{Etot}, which is minimized against the variational
parameters $A,B$ in \eqref{fAB}.  We compare our calculation of the ground
state energy of QFL for quadratic dispersion ($c_4=0$) against the result from
\Rf{TC8905} in Fig. \ref{energyPlotc2}.  The two results agree within $0.2\%$ of the Coulomb
energy, which is sufficient for us to determine the phase diagram for
the chiral superconducting states.

\begin{figure*}[t]
\centering
\begin{subfigure}[t]{0.66\columnwidth}
\centering
\includegraphics[width=\columnwidth]{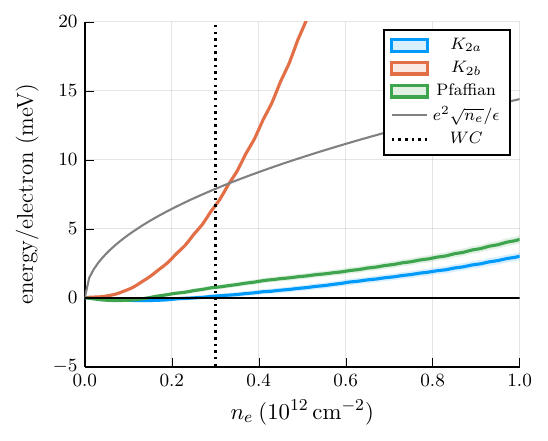}
\caption{$\epsilon = 10$, $c_4 = 549$ m$eV$ n$m^4$}
\end{subfigure}%
\hfill
\begin{subfigure}[t]{0.66\columnwidth}
\centering
\includegraphics[width=\columnwidth]{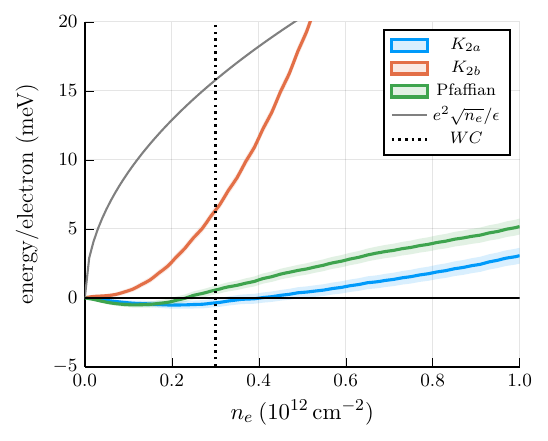}
\caption{$\epsilon = 5$, $c_4 = 549$ m$eV$ n$m^4$}
\end{subfigure}%
\hfill
\begin{subfigure}[t]{0.66\columnwidth}
\centering
\includegraphics[width=\columnwidth]{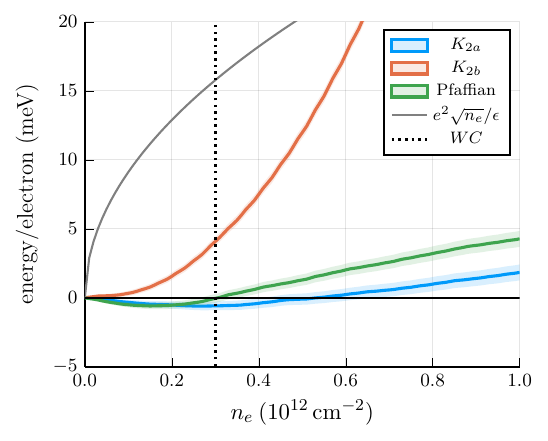}
\caption{$\epsilon = 5$, $c_4 = 366$ m$eV$ n$m^4$}
\end{subfigure}
\caption{Ground state energy per electron for $K_{2a}$, $K_{2b}$, and Pfaffian
superconductors, minus that of QFL,
along the Fermi-surface transition line (the slanted line in
Fig. \ref{fig:phase}). The colored region is the approximate range of error.}
\label{energyPlot}
\end{figure*}

For the quadratic dispersion, the transition to chiral superconductor at low
density remains intact, at densities $n_e =  (0.4\sim 0.7)\times 10^{12}
cm^{-2}$.  As a reference, experimentally\cite{HJ240815233,LJ230917436}, chiral
superconductivity was observed around $n_e = (0.2 \sim 0.7)  \times
10^{12}$cm$^{-2}$, while the Wigner crystal was observed below $n_e < 
(0.2 \sim 0.4) \times 10^{12}$cm$^{-2}$.

We can see the importance of the optimization of the MC variables,
since the Fermi liquid energy improves significantly compared to the ordinary
quadratic dispersive electron gas. Intuitively, the QFL has many
low energy excitations and is therefore well-suited to deform into a lower
energy groundstate.

\section{The Phase Diagram}

From the dispersion relation given in equation~\eqref{eq:disp}, we can set $c_2$ and $c_4$ to numerical values extracted from \Rf{HJ240815233} for
four-layer systems. For this section, we will assume $c_4$ to be a constant, and
consider the possible values to be $c_4=549$ meV-$\mathrm{nm}^4$ and $c_4=366$ meV-$\mathrm{nm}^4$, representing the approximate range of $c_4$.

We identify the region of electron density and $c_2$ where the chiral
superconducting states or QFL are energetically favorable, which leads to the
phase diagram Fig. \ref{fig:phase}.  This indeed shows that as a
positive quadratic term ($c_2>0$) is added, the chiral superconducting states
become gradually unfavorable, appearing at lower electron densities.  If
we add a negative quadratic term ($c_2<0$), the chiral superconducting states
become more favorable.

However, $c_2$ cannot be too negative, since a very negative $c_2$ will cause a
Fermi surface transition where a hole pocket starts to appear near $k=0$.  The
critical value of $c_2$ is given by (see Fig. \ref{FStrans})
\begin{equation}
c_2 k_F^2 + c_4 k_F^4 = 0; \quad c_2 = - c_4 k_F^2 =  - 4\pi n_e c_4.
\end{equation}
The slanted line in Fig. \ref{fig:phase} is the Fermi surface transition line,
above which a hole pocket will be developed and our calculations become invalid
since we did not consider the QFL with a hole pocket. Thus, we should ignore the
phase diagram above the slanted line. In this region, the hole-like Fermi surface will further lower the Fermi liquid energy.

In Fig. \ref{energyPlot}, we plot the ground state energy per electron for
$K_{2a}$, $K_{2b}$, and Pfaffian superconductors (relative to the energy per electron of the QFL) along
the Fermi surface transition line.  We see that the superconducting
condensation energy (per electron) is about $1$m$eV$ or 10K.  In comparison,
the superconducting transition temperature is about $0.3$K.

\subsection{Magnetic Field}
Another knob we can tune is the external magnetic field. The effect of the magnetic field regarding the energetics is given as the Zeeman term $ \pm \mu_B B$ depending on the spin orientation. 

\begin{figure*}[t]
\begin{subfigure}[t]{0.66\columnwidth}
\centering
\includegraphics[width=2.3in]{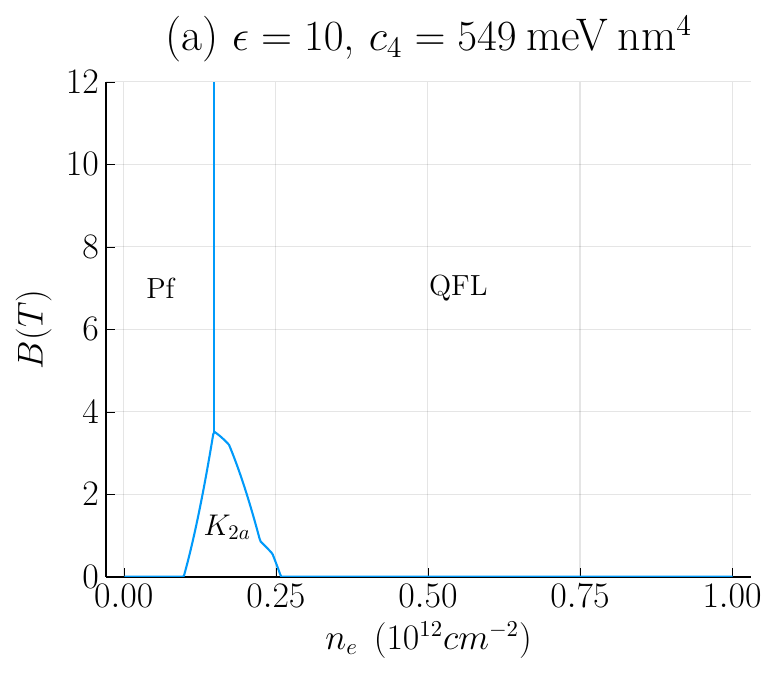}\
\end{subfigure}
\begin{subfigure}[t]{0.66\columnwidth}
\centering
\includegraphics[width=2.3in]{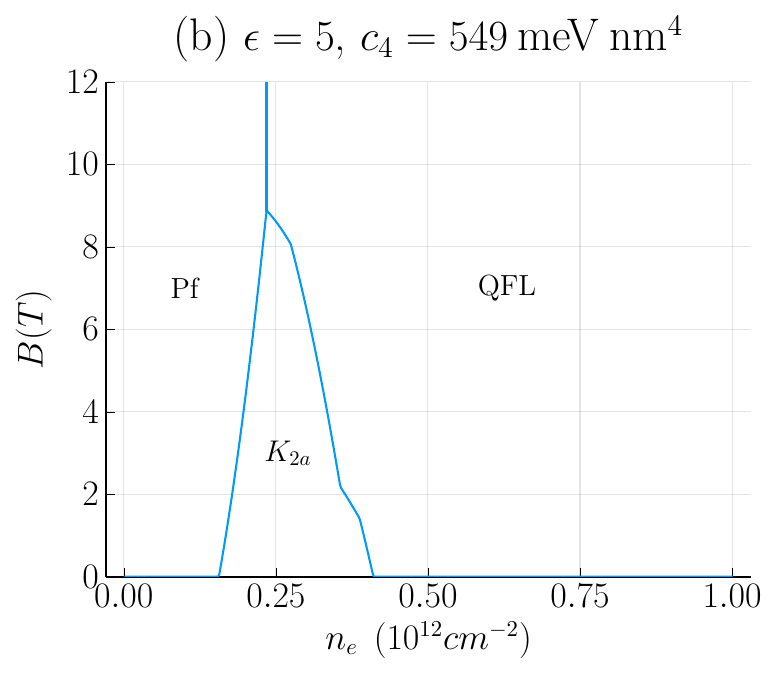}\
\end{subfigure}
\begin{subfigure}[t]{0.66\columnwidth}
\centering
\includegraphics[width=2.3in]{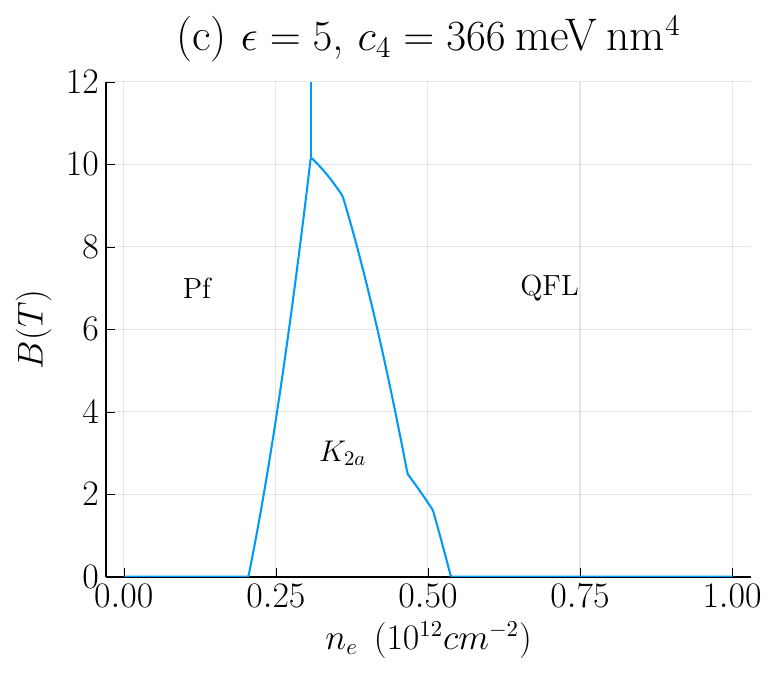}
\end{subfigure}
\caption{The phase diagram for electron dispersion $E_k = c_2 k^2 + c_4 k^4$
along the Fermi surface transition line (\ie for when $c_2 = - 4\pi c_4 n_e$).
The horizontal axis is $n_e \in [0,10^{12} cm^{-2}]$.  The vertical axis is
magnetic field $B$ in teslas.
}
\label{fig:zeeman}
\end{figure*}

For the quarter Fermi liquid and the single-species Pfaffian superconductor, when the external magnetic field is applied, the spin will be polarized to the direction of the magnetic field, and the QFL's energy will be lowered by $ \pm \mu_B B$ per electron. Therefore, the energy difference does not change between the Pfaffian superconductor and the Fermi liquid at this order. At Zeeman order, which we plot in Figure~\ref{fig:zeeman}, if the Pfaffian state is energetically preferable, it will remain stable regardless of the magnetic field. In real materials, the superconductor will be overtaken by the Fermi liquid state by second-order effects. Since this happens at second-order, we can  reasonably conclude that it will occur at high external magnetic fields.

In the case of two-species superconductors, the magnetic field will not change the energy, as half of the
electrons will gain $ \pm \mu_B B$ energy and the other half will lose $ \pm
\mu_B B$, leveling the first-order energy gain. Therefore, the effect of the Zeeman term is to
lower the QFL and Pfaffian energy relative to the two-species chiral
superconducting states. %, and we naturally expect when the magnetic field becomes
%strong enough, the QFL will win over all chiral superconducting states.
Figure~\ref{fig:zeeman} plots the phase diagram along the Fermi surface
transition line between density and magnetic field for various values of the
permittivity and $c_4$.

We observe that, in some cases, the two-species chiral superconducting states can also remain robust for fairly strong magnetic fields along the Fermi surface transition line (where the superconductivity is observed experimentally).  This agrees well with the experimental observations that the superconducting regime does not shrink very much for external fields up to $5$T.  So the current experiments may not totally rule out the possibility of spin-unpolarized chiral superconducting states, such as the $K_{2a}$ state.

\section{Conclusion and Outlook}

We have performed a quantum Monte Carlo optimization of the newly proposed
chiral superconducting states, whose ans{\"a}tze for the two-species case are given as Laughlin-type wavefunctions and a modified Jastrow factor for the Pfaffian wavefunction. In addition to that, we have performed equivalent optimization of the QFL energy. While the results for the quadratic dispersive Fermi liquid are known, we have performed the QMC optimization of the quartic dispersive Fermi liquid using a Slater-Jastrow form wavefunction.

After optimizing all proposed Monte Carlo parameters, we find that the chiral superconducting states can in general win over the QFL at high enough densities to be relevant to experimental systems.  In particular, even a spin unpolarized $K_{2a}$ superconductor (which is in the same phase as spin-triplet $p+\ii p$ BCS superconductor) alongside the spin-polarized Pfaffian superconductor can have a lower ground state energy than that of the QFL, at densities between the Wigner crystal phase and the QFL phase.  The superconducting condensation energy (per electron) is about 5\% of the Coulomb energy $\sqrt{n_e}e^2/\eps$ (with is about 1 meV for four layer samples).

From our studies, we find that of the two types of chiral superconductors which win against the QFL, the Pfaffian superconductor is more optimal at lower densities than $K_{2a}$. However, overall, the difference of energies between the Pfaffian and the $K_{2a}$ states is very small, meaning that both states could still emerge as the ground state in real systems at higher/lower densities than suggested.

\begin{figure}[t]
	\centering
	\includegraphics[width=1.5in]{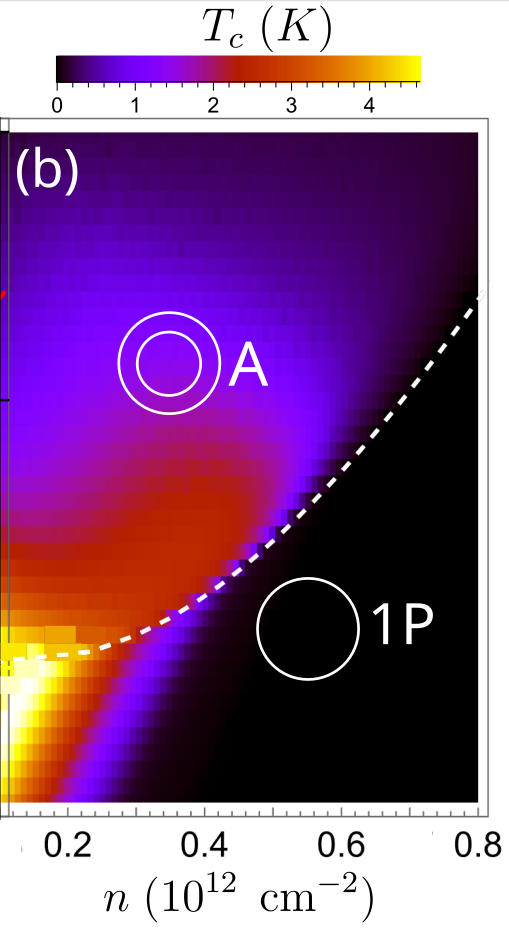}
	\caption{The phase diagram proposed in \Rf{GF240913829} for fully
spin-valley polarized BCS superconductors.  The dashed-line is the
Fermi-surface transition line (see Fig. \ref{FStrans}) For ring-like Fermi
surface above the dashed-line, the superconductor is non-topological with no
chiral edge state.  For disk-like Fermi surface below the dashed-line, the
superconductor is topological with chiral edge state of a single Majorana
fermion, as well as a Majorana zero mode in a magnetic vertex.  Wigner crystal
appears below electron density $\sim 0.2 - 0.4 \times 10^{12} cm^{-2}$. }
\label{SCphaseFu}
\end{figure}

A few interesting future directions are the following. First, one can include the effect of hole pocket, which appears at high displacement fields where the Fermi surface undergoes a Lifshitz transition. Especially, it would be interesting to see that if the superconductivity which emerges at different sides of the Lifshitz transformation could be of different origin. For example, in the phase diagram suggested by \Rf{GF240913829}, a BCS-like superconductor was predicted above the Fermi transition line (see dashed line of Fig.~\ref{SCphaseFu}.)

Another direction is to incorporate the effect of Berry curvature, which will translate into the electrons having a smeared position profile which breaks time-reversal symmetry, possibly favoring one chirality over the other. The energetics when Berry curvature is involved is an open question, and similar questions have been asked for the superconductivity driven in a pairing
context~\cite{JL241109664,MHD250305697}. Considering the Berry curvature will most likely lower the energy of chiral superconducting states, to progress further, it
will be important to include this effect. Qualitatively, it is reasonable to expect that the Berry curvature effects will lower the energy of $K_{2b}$ state most as its orbital angular momentum is the highest. $K_{2a}$ and Pfaffian states will also benefit from this as they are chiral states but less compared to $K_{2b}$. The Fermi liquid will be improved the least, as it does not involve macroscopic orbital angular momentum.

To complete the phase diagram, including the energetics of the Wigner crystal also will be crucial. So far in this paper, we focused on if the chiral superconducting states can energetically be favorable against the quarter Fermi liquid only, as this phase boundary will in general be more delicate due to the optimized kinetic energy of the Fermi liquid phase. The energetics of the Wigner crystal could also elucidate which of the chiral superconducting state options in the end will be most likely to appear.

Since the Wigner crystal will be even more potential-energy optimized than any other competing states, it will also be interesting to have a systematic method of constructing analytical wavefunctions which describe the chiral superconductors, preferably without too many variational parameters. This could further improve what we have designed for the single-species chiral supercondcutor and could be generalized to two-species cases as well. The benefit of this approach is that the kinetic energy terms $\langle k^2 \rangle$ and $\langle k^4 \rangle$ could be readily calculated compared to more a complex form of wavefunctions. Such wavefunctions could also improve our understanding of what properties of the wavefunction can be optimized in a more intuitive way.

%An interesting future direction is to include the effect of hole pocket, which experimental evidence points to at high displacement fields. This is also linked to the more general dispersion relation where $c_2$ is sufficiently negative that the Fermi surface undergoes a Lifshitz transformation; in this regime, the appropriate form of the first-quantized wavefunction is currently unknown. 

%As mentioned above, another direction is to incorporate the effect of Berry curvature, which will translate into the electrons having a smeared position profile which breaks time-reversal symmetry, possibly favoring one chirality over the other. While our work is a standalone work which can be applied to general flat band systems without Berry curvature beyond graphene systems, the change in energetics when the Berry curvature is applied could have an important role in rhombohedral-stacked graphene systems. Indeed, similar question has been asked for the superconductivity driven by pairing context~\cite{JL241109664,MHD250305697}.

The authors thank David Ceperley, Liang Fu, Leyna Shackleton, Max Geier, and
Aidan Reddy for useful discussions on the Monte Carlo calculations
of the Fermi liquid, and Zhengyan Darius Shi regarding the variational
parameters of the chiral superconductors. We also thank Long Ju and Tonghang
Han for their experimental insights. This work was partially supported by the
NSF grant DMR-2022428 (MLK, XGW) and by the Simons Collaboration on Ultra-Quantum Matter, which is a grant from the Simons Foundation (651446, XGW).  AT was supported by NSF GRFP grant 2141064.  Some of the numerical calculations were done on the subMIT HPC cluster at MIT, and some were done using services provided by the OSG Consortium, which is supported by the National Science Foundation awards 2030508 and 2323298.

\appendix

\begingroup
\allowdisplaybreaks
\section{Slater-Jastrow Wavefunction for the quartic dispersive Fermi liquid}
\label{sec:appA}
While the result of the quadratic dispersive electron gas is well known, there has not been an attempt to extract the energy expectation value of the quartic dispersive Fermi liquid, namely with the Hamiltonian form of
\begin{equation}
H = c_4 k^4 + \sum_{i < j} \frac{e^2}{ \lvert r_i -r_j \rvert}.
\end{equation}
In this appendix, we explain the Monte Carlo method of extracting the expectation value of this Hamiltonian. In real space, the Hamiltonian can be written as
\begin{equation}
H = c_4 \nabla^4  + \sum_{i < j} \frac{e^2}{ \lvert r_i -r_j \rvert}.
\end{equation}
where $\nabla^4$ is the biharmonic operator. For variational Monte Carlo, we can sample the energy expectation value by computing $\langle H \rangle$ as
\begin{align}
\langle H \rangle &= \frac{\int \prod_i^N d^2 r_i \psi^* H \psi}{\int \prod_i^N d^2 r_i \psi^* \psi} 
\\
&= \frac{\int \prod_i^N d^2 r_i \psi^* \psi \frac{H \psi}{\psi}}{\int \prod_i^N d^2 r_i \psi^* \psi} 
= \left\langle \frac{H \psi}{\psi}\right\rangle_{MC},
\nonumber 
\end{align}
where $\psi$ is the many-body wavefunction. Since $\psi^* \psi$ represents the probability distribution function of the given wavefunction, this can be calculated by sampling the often called \textit{local energy} $\frac{H \psi}{\psi}$ with the Monte Carlo algorithm. 

For the simulations, we will set our electron number $N=69$ at $n_e = 1$ units, with the electrons subject to periodic boundary conditions to attenuate boundary effects. Our trial function is given as a Slater-Jastrow format, 
\begin{equation}
\psi(R)= D(R) \exp (J),
\end{equation}
where $R$ is the collective coordinates of $N$ particles $r_1, r_2 \cdots, r_N$, $D(R)$ is the Slater determinant, and $J$ is the Jastrow factor which depends on the relative distance between two particles, given as
\begin{equation}
J = \sum_{i < j} f(\lvert r_i - r_j \rvert).
\end{equation}
The optimal choice of $f$ is unexplored in previous literature. We propose an ansatz
\begin{equation}
f(r) = A e^{-\frac{r}{B}}\left(1+ \frac{r}{B} + \frac{r^2}{2B^2}\right).
\end{equation}
This potential has the advantage of $f'(0) = f''(0) = 0$, which helps moderating the divergences of the kinetic energy when two particles are close to each other. In general, when only regarding the correlation energy, the exact form of the Jastrow factor is not sensitive to a specific choice~\cite{C7818}. We retrieve the Hartree-Fock limit when $A \rightarrow 0$, giving a powerful sanity check to our calculations. Note that while the Jastrow factor is short-ranged, at large enough B there is still the need to consider supercell summations of the Jastrow factor. We use 9 supercell images, which turns out to give enough range of Monte Carlo parameters to find the energy minimum.

Going back to the sampling of local energy, the potential part is $\left\langle \frac{H \psi}{\psi}\right\rangle_{MC} = \left\langle \sum_{i<j} \frac{1}{\lvert r_i - r_j \rvert} \right \rangle$. Unfortunately, in 2D this potential remains long-ranged, and also a uniform positive background has to be added to maintain charge neutrality of the overall system. We can obtain the effective resummation of all the images with subtracting the background via the Ewald summation as done by many previous works~\cite{ND571, BST6645, C7818} dealing with periodic boundary conditions contains a general discussion of Ewald summation; for the application to the electron gas see Ref.~\cite{C7818}. We follow the full derivation outlined in Appendix A of Ref.~\cite{GNZF250205383}. To summarize the results, the total unregulated potential energy can be written as
\begin{equation}
V = \frac{1}{2} \sum_{i < j, \mathbf{a}} \frac{1}{\lvert \mathbf{r}_i - \mathbf{r}_j + \mathbf{a} \rvert} + \sum_{i, \mathbf{a} \neq 0} \frac{1}{\lvert \mathbf{a} \rvert} 
\end{equation}
where $\vec{a}$ is the lattice vector given as the lengths of the two axes of the torus. This value obviously diverges, which can be controlled by imposing a uniform charge background \textit{a la} Jellium; the uniform background will correspond to the $q=0$ contribution to the Fourier transformed potential energy. The long-rangedness of the potential can be confirmed by the Fourier transform, where
\begin{equation}
\phi_{\vec{a}, i} = \int d\mathbf{r} \frac{e^{i \mathbf{q} \cdot \mathbf{r}}}{\lvert \mathbf{r} - \mathbf{r}_i + \mathbf{a} \rvert} = 2\pi \frac{e^{-i\mathbf{q} \cdot (\mathbf{r}_i + \mathbf{a})}}{q}
\end{equation}

The key idea is that the principal contributions to the energy can be mostly resummed by adding up the short-range Coulomb energy effects by summing over small values of $\mathbf{a}$ on one end, and then add up the short-range terms of the \textit{reciprocal} lattice vectors $\mathbf{G}$, which will contain most of the long-range divergences. The total regulated potential energy can be written as
\begin{equation}
V=\frac{1}{2} \sum_{i<j} \phi_{i} (\mathbf{r}_j) + \frac{1}{2} \sum_i \xi_M
\end{equation}
where
\begin{multline}
\phi_{i} (\mathbf{r}_j) = \sum_{\mathbf{a}} \frac{1}{\lvert \mathbf{r}_j - \mathbf{r}_i + \mathbf{a} \rvert } \textrm{Erfc} \left( \frac{\lvert \mathbf{r}_j - \mathbf{r}_i + \mathbf{a} \rvert }{2\eta} \right) \\
-\frac{2\pi}{A} \frac{2\eta}{\sqrt{\pi}}+ \frac{2\pi}{A} \sum_{\mathbf{G} \neq 0 } e^{i \mathbf{G} \cdot (\mathbf{r}_j- \mathbf{r}_i)} \frac{\textrm{Erfc} (\eta \lvert \mathbf{G} \rvert )}{\lvert \mathbf{G} \rvert}
\end{multline}
and
\begin{multline}
\xi_M = \sum_{\mathbf{a}} \frac{1}{\lvert \mathbf{a} \rvert } \textrm{Erfc} \left( \frac{\lvert  \mathbf{a} \rvert }{2\eta} \right) -\frac{2\pi}{A} \frac{2\eta}{\sqrt{\pi}} \\
+ \frac{2\pi}{A} \sum_{\mathbf{G} \neq 0 } \frac{\textrm{Erfc} (\eta \lvert \mathbf{G} \rvert )}{\lvert \mathbf{G} \rvert} -\frac{1}{\eta \sqrt{\pi}}
\end{multline}
where $\eta$ is a cutoff parameter which we can adjust. For sufficient number of summations of real lattice vectors $\mathbf{a}$ and reciprocal lattice vectors $\mathbf{G}$, $V$ should converge to the same value at a wide enough range of $\eta$. Empirically, taking 30 real and reciprocal lattice vectors respectively is enough to guarantee good convergence, of which we can check the soundness of the value calculated by comparing it to the Hartree-Fock result.

For the kinetic part, the principal difficulty is in analytically obtaining $\frac{\nabla^4 e^J}{e^J}$; until $\frac{\nabla^2 e^J}{e^J}$ however, the analytical form is well known. We can avoid this problem by observing that periodic boundary condition puts the system on a torus, which conveniently does not have a boundary. The key idea therefore is performing integration by parts without worrying about the finite-size effects here. The expectation value of the kinetic energy can be written
\begin{align}
\langle T \rangle &= \frac{\int \prod_i^N d^2 r_i \psi^* \nabla^4 \psi}{\int \prod_i^N d^2 r_i \psi^* \psi} \\
&= \frac{\int \prod_i^N d^2 r_i \nabla^2 \psi^* \nabla^2 \psi}{\int \prod_i^N d^2 r_i \psi^* \psi} \\
&= \frac{\int \prod_i^N d^2 r_i \psi^* \psi  \left\lvert \frac{\nabla^2 \psi}{\psi} \right \rvert^2}{\int \prod_i^N d^2 r_i \psi^* \psi} \\
&= \left\langle \left\lvert \frac{\nabla^2 \psi}{\psi} \right \rvert^2\right\rangle_{MC}.
\end{align}
Therefore, sampling the kinetic part of the local energy is of the same computational complexity as a regular quadratic dispersive system. The $A=0$ Hartree-Fock case and comparing the energies to the analytical values, which is $\langle k^2 \rangle = 2\pi$ and $\langle k^4 \rangle = \frac{16\pi^2}{3}$ at $n_e = 1$ units; the result agrees well with smaller than $0.2\%$ error. A further check is comparing the sampling at the quadratic order between $\left\langle \left\lvert \frac{\nabla \psi}{\psi} \right \rvert^2\right\rangle_{MC}$ and $\left \langle \frac{\nabla^2 \psi}{\psi} \right \rangle_{MC}$. The results here also give a smaller than $0.5\%$ error for all MC parameter values sampled, further corroborating the robustness of taking the alternative sample.

We perform a two-parameter optimization of the Jastrow factor: $A$ is the overall amplitude, and $B$ is the approximate range where the Jastrow correlation will be significant. For both, we take the $n_e = 1$ units; as we change the density, the $A$ and $B$ parameters will then be automatically tuned. We perform VMC sampling for $A =  0.5, 1, 1.5, 2, 2.5, 3, 4, 5, 6, 7, 8, 10, 12, 15$ and take $B = 0.1 $ to $1.0$ in $0.1$ intervals. For each $A$ and $B$ values, we take 20 independent MC runs, with each sample containing $5\times 10^5$ data points.

A point to confirm is if our Jastrow factor captured a good portion of the total correlation energy possible. While some discussion of this was done at the main text, it is also useful to directly see the difference in correlation energy for the quadratic dispersive electron gas where a good fit of the energy is given as the Gaskell form~\cite{G6177, G6280} at $\epsilon = 5$, which is plotted in Figure~\ref{fig:quadcheck}. 

\begin{figure}[t]
	\begin{subfigure}[t]{0.48\columnwidth}
		\centering
		\includegraphics[width=1.7in]{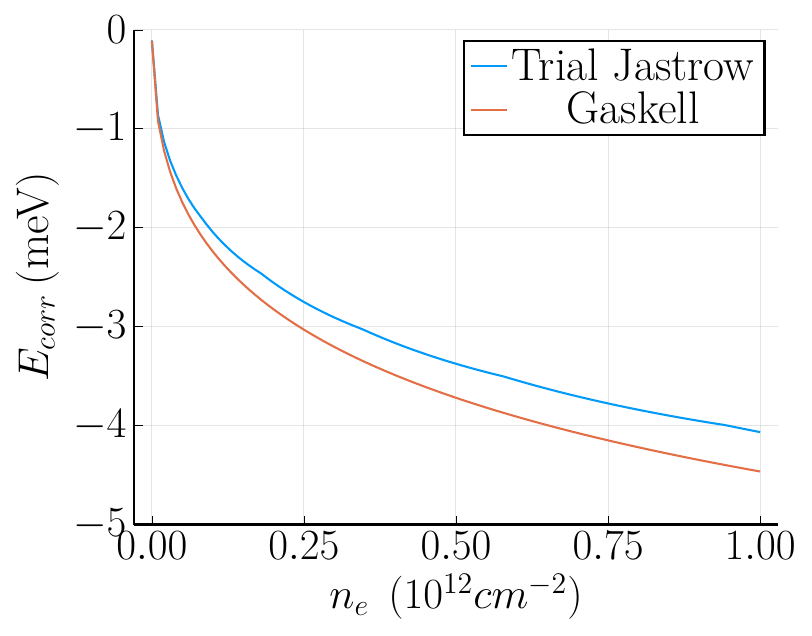}\
	\end{subfigure}
	\begin{subfigure}[t]{0.48\columnwidth}
		\centering
		\includegraphics[width=1.7in]{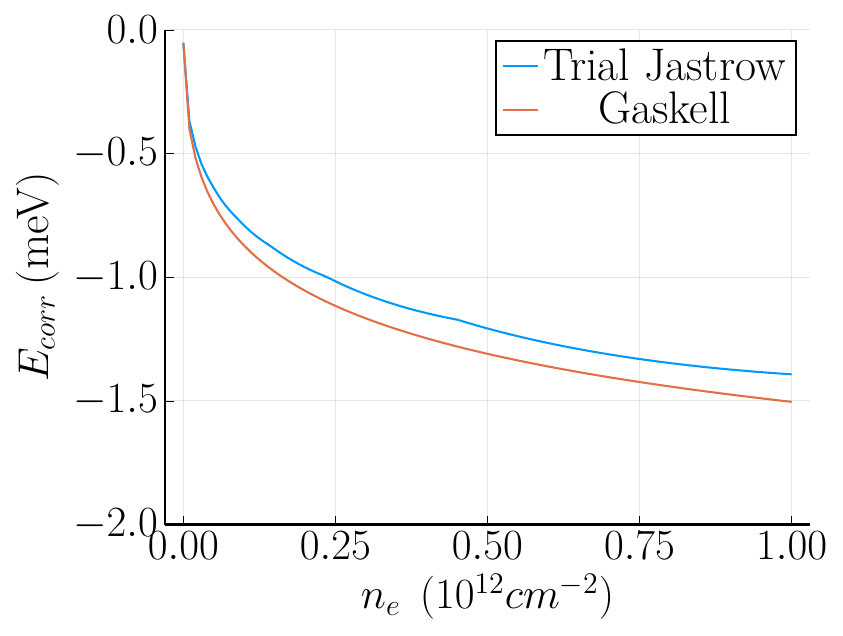}\
	\end{subfigure}
	\caption{Comparison of the correlation energy for our trial Jastrow factor versus the Gaskell form when (left) $\epsilon = 5$ and (right) $\epsilon = 10$. We observe that our class of Jastrow factor captures most of the correlation energy for quadratic dispersive fully polarized Fermi gas. For the parameters, we substitute the parameter extraction for $c_4$ directly to $c_2$, which gives $c_2 = 91 \textrm{meV} \: \textrm{nm}^2$.}
	\label{fig:quadcheck}
\end{figure}

The difference between the well-optimized Gaskell form to our trial Jastrow is less than 10\% of the correlation energy, which at phase transition region around $5.0 \times 10^{11} \textrm{cm}^{-2}$. Considering that our chiral superconducting state wins over the Fermi liquid around with 1 meV energy difference at this ansatz, we can conclude that even at the ideal case the chiral superconductor will win over the Fermi liquid at reasonable densities. The difference in energy is even smaller for $\epsilon = 10$, as shown on the right figure of Figure~\ref{fig:quadcheck}.

\section{Kinetic and Potential Energy of Two-Species Chiral Superconductors}
\label{sec:appB}

\subsection{Kinetic Energy}
The kinetic energy for the two-species chiral superconductor can be directly sampled using the local energy. As we did for the quarter Fermi liquid in the previous section, we can sample $\langle k^2 \rangle$ and $\langle k^4 \rangle$ from sampling the local energy. For $\langle k^2 \rangle$, we use $-\frac{\nabla^2 \psi}{\psi}$. While we are running our Monte Carlo simulations, the integration is still over the whole $\mathbb{R}^2$, allowing for the integration by parts trick used for the quarter Fermi liquid of sampling $\left\lvert \frac{ \nabla^2 \psi}{\psi} \right\rvert^2$ to still apply. However, since the ansatz wavefunctions do not involve complex functions such as the determinant, it was numerically viable just to sample $\frac{\nabla^4 \psi}{\psi}$ at $N=200$ electrons.

One additional complication is the edge effect, where the kinetic energy for electrons near the edge increases compared to the electrons sufficiently inside the bulk. Therefore, we restrict the sampling of the local energies to electrons only sufficiently inside the bulk. For a general quantum Hall droplet of radius $R$, we only sample local energies for electrons inside $R/3$ from the origin. To check that this is enough, we test for many different system sizes (electron numbers) to verify that the kinetic energy per particle gives identical results. Note that by taking this method, while the total kinetic energy of the electrons are real, if we restrict to electrons within $\frac{1}{3}  R$ from the origin at each uncorrelated sample $-\frac{\nabla^2 \psi}{\psi}$ and $\frac{\nabla^4 \psi}{\psi}$ will be a complex number. The expectation is that after sufficient number of samples, the imaginary part will cancel out and the kinetic energy per particle could be read off only by looking at the real component. Indeed, results show that the after taking $3 \times 10^5$ samples is enough to make the imaginary part negligible. From this procedure, we also can numerically verify that indeed sampling $\left\lvert \frac{ \nabla^2 \psi}{\psi} \right\rvert^2$ and $\frac{\nabla^4 \psi}{\psi}$ retrieves the same result, differing less than $0.2\%$ from each other.

\subsection{The Potential Energy}
The interaction energy between a species-$I_0$ particle and a species-$J_0$ particle is given as:
\begin{align}
U_{I_0J_0}
\int \prod_{I,i} 
d^2 z_i^I V(z^{I_0} - z^{J_0})
|\Psi(z^{I_0}, z^{J_0}, \{z_i^I\})|^2.
\end{align}
The pair correlation function of the two electrons is defined as
\begin{align}
\frac{g_{I_0J_0}(z^{I_0} - z^{J_0})}{(\pi R^2)^2}
= 
\int \prod_{I,i} 
\dd^2 z_i^I 
|\Psi(z^{I_0}, z^{J_0}, \{z_i^I\})|^2
\end{align}
as for the Laughlin-type wavefunction at the thermodynamic limit, the density is constant throughout the droplet of radius $R$. Since $g_{I_0J_0}(z^{I_0} - z^{J_0})$ becomes a constant when $|z^{I_0} -
z^{J_0}|$ is larger than a finite correlation length, and since 
\begin{align}
\int_{\pi R^2} 
\dd^2 z^{I_0}
\dd^2 z^{J_0} \
\frac{g_{I_0J_0}(z^{I_0} - z^{J_0})}{(\pi R^2)^2} =1,
\end{align}
we see that $g_{I_0J_0}(z^{I_0} - z^{J_0}) =1$ when $|z^{I_0} - z^{J_0}|$ is
larger than the correlation length.
We find
\begin{align}
U_{I_0J_0} =
\int_{\pi R^2} 
\dd^2 z^{I_0}
\dd^2 z^{J_0} \  V(z^{I_0} - z^{J_0})
\frac{g_{I_0J_0}(z^{I_0} - z^{J_0})}{(\pi R^2)^2}
\end{align}
For Coulomb interaction, we also need include the background charge:
\begin{align}
U_{I_0J_0} &=
\int_{\pi R^2} 
\dd^2 z^{I_0}
\dd^2 z^{J_0} \  V(z^{I_0} - z^{J_0})
\frac{g_{I_0J_0}(z^{I_0} - z^{J_0})-1}{(\pi R^2)^2}
\nonumber\\
&=
\int
\dd^2 z \ \frac{e^2}{\eps |z|} 
\frac{g_{I_0J_0}(z)-1}{\pi R^2}
\end{align}
From here, the interaction energy per particle is therefore
\begin{align}
E_\text{int} &= \frac1{2N} \sum_{IJ} U_{IJ} N_I N_J
\nonumber\\
&=
\frac12 \sum_{IJ} f_I f_J n_e\int
\dd^2 z \ \frac{e^2}{\eps |z|} (g_{IJ}(z)-1)
\end{align}
The typical separation between
particles is $n_e^{-1/2}$. Rewriting this to extract out the dimensionless factor
\begin{align}
E_\text{int} 
&=
\sum_{IJ} f_I f_J 
\frac{e^2 \sqrt{n_e}}{2\eps}  
\int
\dd^2 z \ \frac{ \sqrt{n_e}}{ |z|} (g_{IJ}(z)-1)
\nonumber\\
&=
\frac{e^2\sqrt{n_e}}{\eps}  
\sum_{IJ} f_I f_J V_{IJ}
=
\frac{e^2\sqrt{n_e}}{\eps} V 
\label{eq:EandV}
\end{align}
where
\begin{align}
V_{IJ} &\equiv \int
\dd^2 z \ \frac{ \sqrt{n_e}}{2|z|} (g_{IJ}(z)-1) .
\nonumber\\
V &\equiv \sum_{IJ} f_I f_J V_{IJ}.
\label{eq:totenergy}
\end{align}
This is the generalization compared to the one-electron species case~\cite{L8395, MH8633}. For this step we have used the fact that $\sqrt{n_e}$ has units of inverse length, so $e^2 \sqrt{n_e}$ has the units of Coulomb interaction.

Given a $K$-matrix, our wavefunction ansatz has a free parameter of deforming $\lvert z_i^I - z_j^J \rvert^{2\alpha}$, which is represented at $\bar{K}$ in our formulation. 

The potential energy almost solely depends on $\bar{K}$. $K$ does play a limited role by setting the fraction of species-$I$ electrons; for two-species case, this is $f_I = (1/2, 1/2)$. We numerically calculate $V_{IJ}$ for all $\bar{K}$ in
\begin{equation}
\bar{K} = \begin{pmatrix}
a & c \\ c & b
\end{pmatrix}
\label{eq:barKrep}
\end{equation} 
format, fixing $f_I = (1/2, 1/2)$ and modifying the magnetic length $l_I$ accordingly to impose the same droplet radius for the two species. Without losing generality, we will assume $a \geq b$, and more specifically $a \geq b \geq c$. 

We have run our variational Monte Carlo routine for $\bar{K} = \begin{pmatrix} a & c \\ c & b \end{pmatrix}$ for all $a, b \geq 1$ and $5 \geq a \geq b \geq c \geq 0$ at $0.1$ intervals for 200 electrons. We ignore the first $3 \times 10^5$ MC steps for the system to equilibrate, and sample until $3 \times 10^6$ steps for a MC configuration. We have sampled from 20 independent MC configurations.

A good sanity check for the potential energy is comparing the numerical results with the already known single-species case~\cite{L8395, MH8633}. On our formalism, this corresponds to the case when all inter- and intra-species interactions are the same, which is $\begin{pmatrix} a & a \\ a & a \end{pmatrix}$.

\subsection{Alternative Method for Kinetic Energy Sampling}

The kinetic energy of the ansatz wavefunction can be calculated in a different way, naley via computing its Green's functions, which we will quote the result~\cite{KW240918067}:
\begin{align}
&\ \ \ \
G_{I_0}( z, z^*, \tl z, \tl z^*) 
\\
& =
C\big(1
+ g_2 (\tl z^*-z^*) (z-\tl z)
+ g_4 (\tl z^*-z^*)^2 (z-\tl z)^2\big)
\nonumber \\
&\ \ \ \
\ee^{\sum_{IJ} 
\pi n_J
K_{I_0J}^+
z \tl z^* 
-\frac{ |z|^2}{4l_{I_0}^2}}
\ee^{\sum_{IJ}  
\pi n_J
K_{I_0J}^-
\tl z z^* 
-\frac{ |\tl z|^2}{4l_{I_0}^2}}
\nonumber 
\label{eq:green}
\end{align}
Note that the density profile of species-$I_0$ particle is the special case $G_{I_0}( z, z^*,  z,  z^*) =\rho_{I_0}(z)$. For Laughlin-type wavefunction, we know that this quantity should be a uniform throughout the droplet which is a disk. Therefore, $G_{I_0}( z, z^*,  z,  z^*)= \frac{1}{\pi R^2}$ constant.

%Inspired by the recent experiments in tetralayer graphene~\cite{HJ240815233,LJ230917436}, let us now assume that the particle is subject to quadratic energy dispersion and/or a quartic energy dispersion, with the latter being a model for the flat bands. 

Starting with the (purely) quadratic dispersion, the corresponding kinetic energy operator is $-\prt_x^2-\prt_y^2 = -4
\prt_{z^*} \prt_z$. The average kinetic energy is then given by
\begin{align}
&\ \ \ \ -4\int_{\pi R^2} \dd^2 z\; \prt_{ z^*} \prt_z
G_{I_0}( z, z^*, \tl z, \tl z^*) \mid_{z=\tl z}
\nonumber \\
&= -4\int_{\pi R^2} \frac{\dd^2 z}{\pi R^2}\
|z|^2 \Big(\sum_{IJ} \pi n_J K_{I_0J}^+ 
-\frac{1}{4l_{I_0}^2}\Big)
\nonumber\\
&
\ \ \ \ \ \ \ \ \ \ \ \ \ 
\Big(\sum_{IJ} \pi n_J K_{I_0J}^- 
-\frac{1}{4l_{I_0}^2}\Big)
-\frac{1}{4l_{I_0}^2}
- g_{2, I_0}
\nonumber \\
&=
\frac{1}{l_{I_0}^2} +4 g_{2, I_0} = \sum_I 2\pi n_I \bar K_{II_0} +4g_{2, I_0}
\end{align}
In the above calculation, we observe that unless $\frac{1}{l_I^2}$ satisfies
\begin{align}
\label{KplI}
\sum_{IJ} K_{IJ}^+ \pi n_J = \sum_{IJ} K_{IJ}^- \pi n_J=
\frac14 \frac{1}{l_{I}^2} 
\end{align}
the average kinetic energy of a single particle diverges at the thermodynamic limit.  This is equivalent to the condition we found earlier based on angular momentum.

In terms of $K$ and $\bar{K}$, these to conditions can be rewritten as first,
\begin{align}
\sum_{J} K_{IJ} n_J = 0,\ \ \
\text{where} \ \ K = K^+ -  K^-. 
\end{align}
giving the constraint that $n_J$ has to be in the null space of $K$. To give a reasonable physical situation, all elements has to be $n_J >0$.

The second equation is 
\begin{align}
\sum_{J} \bar K_{IJ} n_J = \frac{1}{2\pi l_I^2},
\end{align}
which allows us to calculate the effective magnetic length of each species of electrons. Provided $n_I$ are all positive, the magnetic lengths will also remain positive. In other words, this condition ensures that the droplet radius $R_I$ for all species $I$ will be the same.

With these two restrictions in mind, let us calculate the kinetic energy of the species-$I_0$ particle when it has a quartic dispersion: 
$(\prt_x^2+\prt_y^2)^2 =16 
\prt_{z^*}^2 \prt_z^2$, therefore the average kinetic energy
is given by
\begin{align}
&\ \ \ \ 16\int_{\pi R^2} \dd^2 z\; \prt_{ z^*}^2 \prt_z^2
G_{I_0}( z, z^*, \tl z, \tl z^*) \mid_{z=\tl z}
\nonumber \\
&= 16\int_{\pi R^2} \frac{\dd^2 z}{\pi R^2}\
\frac{1}{8l_{I_0}^4} + 4g_{4, I_0}
=
\frac{2}{l_{I_0}^4}+64 g_{4, I_0} 
\nonumber\\
&= 2(\sum_I 2\pi n_I \bar K_{II_0})^2 + 64 g_{4, I_0}
\end{align}
assuming  $\frac{1}{l_I^2}$ satisfy
\eqref{KplI}. The factors $g_2$ and $g_4$ have to be extracted numerically. 

\section{A model of electron distribution in $\mathbf{k}$-space}
\label{sec:appE}

\begin{figure}[tb]
\centering
\includegraphics[width=1.0in]{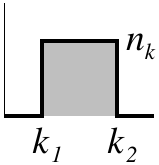}
\caption{A model distribution for
fermion occupation number in each $\v k$-orbital.
} \label{nkdist}
\end{figure}

From the averages $\<k^2\>$, $\<k^4\>$ and the total electron density $n_e$, we
can obtain a model of electron distribution in $\v k$-space.  We assume the
electron occupation per $\mathbf{k}$ orbital is a constant $n_k$ for $k_1 < k <
k_2$, and zero otherwise (see Fig. \ref{nkdist}).  Thus
\begin{align}
n_e &=\int_{k_1}^{k_2} \frac{2\pi k \dd k}{(2\pi)^2} n_k
= \frac{k_2^2-k_1^2}{4\pi} n_k
\nonumber\\
\<k^2\> n_e &= \int_{k_1}^{k_2} k^2 \frac{2\pi k \dd k}{(2\pi)^2} n_k
= \frac{k_2^4-k_1^4}{8\pi} n_k
\nonumber\\
\<k^4\> n_e &= \int_{k_1}^{k_2} k^4 \frac{2\pi k \dd k}{(2\pi)^2} n_k
= \frac{k_2^6-k_1^6}{12\pi} n_k
\end{align}
We see that
\begin{align}
\<k^2\> &= \frac12 (k_2^2+k_1^2), \ \ \ \ \
\<k^4\> = \frac13 (k_2^4+k_1^2k_2^2+k_1^4)
\nonumber\\
\<k^4\> - \<k^2\>^2 &=
\frac{k_2^4+k_1^2k_2^2+k_1^4}{3} -
\frac{k_2^4+2k_1^2k_2^2+k_1^4}{4}
\nonumber\\
&= \frac{1}{12}
(k_2^4-2k_1^2k_2^2+k_1^4) = \frac{(k_2^2-k_1^2)^2}{12}
\end{align}

We find
\begin{align}
n_k &= \frac{4\pi n_e}{k_2^2-k_1^2}
= \frac{2\pi n_e}{\sqrt{3(\<k^4\> - \<k^2\>^2)}}
\end{align}
For $K_{2a}$ and $K_{2b}$ states, $n_k$ must be less then $2$.
For Pfaffian and QFL states, $n_k$ must be less then $1$.
Indeed, our numerical results satisfy this condition.

From
\begin{align}
k_2^2-k_1^2 = \frac{4\pi n_e}{n_k}, \ \ \ \ \ 
k_2^2+k_1^2 = 2\<k^2\>,
\end{align}
we find
\begin{align}
\frac{k_2^2}{4\pi n_e} = \frac{\<k^2\>}{4\pi n_e} +\frac{1}{2n_k}  
,\ \ \ \ \
\frac{k_1^2}{4\pi n_e} = \frac{\<k^2\>}{4\pi n_e} -\frac{1}{2n_k} . 
\end{align}
For QFL, $\frac{k_2^2}{4\pi n_e} = 1$ and $\frac{k_1^2}{4\pi n_e} = 0$.

\bibliography{all, publst}

%\begin{thebibliography}{99}
%\bibitem{laughlin} Laughlin 1988.
%\end{thebibliography}

\end{document}